\documentclass[twocolumn]{aastex631}

\usepackage{CJKutf8}
\usepackage{longtable}
\usepackage{booktabs}
\usepackage{enumitem}
\usepackage{amsmath}
\usepackage{makecell}

\begin{document}

\title{TESS planets in known radial velocity cold Jupiter systems: Hot super Earth occurrence is enhanced by cold Jupiters}

\correspondingauthor{Wei Zhu}
\email{weizhu@tsinghua.edu.cn}

\author[0009-0007-6412-0545]{Quanyi Liu \begin{CJK*}{UTF8}{gbsn}(刘权毅)\end{CJK*}}
\affiliation{Department of Astronomy, Tsinghua University, Beijing 100084, China}

\author[0000-0003-4027-4711]{Wei Zhu \begin{CJK*}{UTF8}{gbsn}(祝伟)\end{CJK*}}
\affiliation{Department of Astronomy, Tsinghua University, Beijing 100084, China}

\author[0000-0002-4503-9705]{Tianjun Gan \begin{CJK*}{UTF8}{gbsn}(干天君)\end{CJK*}}
\affiliation{Department of Astronomy, Westlake University, Hangzhou 310030, Zhejiang Province, China}

\author[0000-0002-8958-0683]{Fei Dai \begin{CJK*}{UTF8}{gbsn}(戴飞)\end{CJK*}}
\affiliation{Institute for Astronomy, University of Hawai‘i, 2680 Woodlawn Drive, Honolulu, HI 96822, USA}

\begin{abstract}
The correlation between inner super-Earths (SEs) and outer cold Jupiters (CJs) provides an important constraint on the formation and dynamical evolution of planetary systems. Previous studies have suggested a positive connection between these two populations, particularly around metal-rich stars, and proposed that nearly all CJ-hosting stars may also harbor inner SEs. In this work, we use TESS transits to investigate the occurrence of hot super-Earths (HSE; 1--4$R_\oplus,\,P<10\,\rm d$) in systems with known CJs detected by radial velocity. Out of a statistical sample of 132 CJ systems, we identify five transiting HSEs around four stars, including one new candidate (TOI-6965.01) around HD 50554. To enable statistical analysis, we first validate the two candidates around HD 50554 using TESS photometry, archival RV measurements, and Gaia astrometry. After accounting for detection sensitivity and geometric transit probability, we find that the presence of CJs enhances the occurrence rate of HSEs by a factor of $6.5^{+3.1}_{-2.3}$ relative to field stars, with the case of no enhancement being ruled out at 99.9\% confidence level. Taking into account the average multiplicity of HSEs, we find that about 91\% of CJ systems host at least one HSE. Our results provide strong supporting evidence for a positive HSE–CJ correlation. We also briefly explore the correlation around metal-poor hosts and for specific sub-populations (e.g., warm super Earths or cold super Jupiters).
\end{abstract}


\section{Introduction} \label{sec:intro}
The connection between close-in small planets—most notably super-Earths (SEs), defined as planets with masses or radii between those of Earth and Neptune—and cold Jupiters (CJs), which are gas giants located beyond $\sim1\,$au, has attracted increasing attention in exoplanet studies. This connection is closely related to the search for true Solar System analogs \citep[e.g.,][]{Zhu:2021, Fernandes:2025}. It also provides important constraints on the physical processes that govern the formation and evolution of planetary systems \citep[e.g.,][]{Bitsch:2020, Schlecker:2021, Guo:2023, Best:2024, Huang:2025, Danti:2025}.

From the observational perspective, \citet{Zhu:2018a} and \citet{Bryan:2019} first reported that the occurrence rate of CJs in systems hosting SEs, denoted as $P(\rm CJ|SE)$, is about 30\%, which is about three times higher than that in field stars without regard to inner small planets. This suggests a positive correlation between the two planet populations. Following these early works, some later studies also found a positive correlation \citep[e.g.][]{Herman:2019, Rosenthal:2022, Zandt:2025}, while some others reported an indifferent or even negative correlation {\citep[e.g.][]{Bonomo:2023, Barbato:2018}}. Such discrepancies can be largely reconciled when considering the metallicity and mass dependences of the cold Jupiter occurrence rate \citep{Zhu:2024, Bryan:2025}, and the correlation appears to be significant at least around metal-rich Sun-like stars \citep{Bryan:2024, Bonomo:2025}. For other stellar types and bulk metallicities, the available dataset is not large enough to tell the difference between positive and no correlation.

For practical reasons, the majority of the existing studies have focused on measuring $P{\rm CJ|SE})$, whereas the inverted conditional rate, $P({\rm SE|CJ})$, namely the occurrence of SEs in systems hosting CJs, has rarely been directly constrained {(though see attempts of \citealt{Barbato:2018, Pinamonti:2023})}. This inverted conditional rate is more directly related to theories that attempt to explain the SE--CJ connection \citep[e.g.,][]{Schlecker:2021, Bitsch:2023, Danti:2025}.
By applying Bayes’ theorem, \citet{Zhu:2018a} and \citet{Bryan:2019} inferred that nearly all CJs should have at least one inner SE companion. However, this predicted high conditional probability carries substantial uncertainty due to Bayesian error propagation. Two studies have attempted to constrain $P({\rm SE|CJ})$ directly. 
In searching for long-period ($>2\,$yr) transiting giant planet candidates in the \textit{Kepler} data, \citet{Herman:2019} found five out of their 13 long-period giant candidates hosted inner small companions. With the assumption that the inner and the outer systems are not highly misaligned ($\lesssim 4^\circ$), they reported that $\sim90\%$ of systems hosting long-period giants also contain inner small planets \citep[see also][]{Masuda:2020}. In addition to the unknown mutual inclinations, another issue with the \citet{Herman:2019} approach is that not all of their giant planets in size are actually giant planets in mass, as some other similar objects have been known to have rather small masses \citep[e.g., HIP 41378~f,][]{Santerne:2019}.
Another direct measurement of the inverted conditional rate is by \citet{Rosenthal:2022}. Using a radial-velocity (RV) sample from the California Legacy Survey (CLS, \citealt{Rosenthal:2021}), \citet{Rosenthal:2022} found that the occurrence rate of inner ($<1\,$au) low-mass (2--30$\,M_\oplus$) planets increases from 27\% to 42\% in the presence of an outer (0.23--10\,au) massive ($>0.1\,M_{\rm J}$) planet. Their definitions of inner and outer planets differ from the commonly adopted ones, making it hard to have a direct comparison. Additionally, the use of a pure (and possibly biased) RV sample limits the sensitivities to only relatively massive planets, whereas the lower-mass but more abundant planets cannot be probed.


RV surveys have undoubtedly been very successful in delivering hundreds of exoplanet detections and constraining the distribution of exoplanets, especially the relatively massive ones \citep[e.g.,][]{Cumming:2008, Mayor:2011, Howard:2011}. However, the current RV surveys are typically insufficient in detecting Earth-mass or even super Earth-mass planets at relatively wide orbits. For example, a $5\,M_\oplus$ planet at 50-day orbit around a Solar-mass star only produces a RV semi-amplitude of $\sim1\,$m/s, which is detectable only by hyper stable instruments on quiet stars with aggressive observing strategies. Detection of low-mass planets in the presence of massive giant planets is even harder \citep[e.g.,][]{AngladaEscude:2010}.
The large-amplitude, long-period RV signal induced by the outer giant can dominate the RV time series, making it challenging to disentangle the low-amplitude signals of inner small planets. In comparison, space-based transit surveys have been proven to be more efficient in detecting small planets, and these detections are largely unaffected by the presence of additional bodies in the system \citep[e.g.,][]{Lissauer:2011, Fabrycky:2014}.

The Transiting Exoplanet Survey Satellite \citep[TESS,][]{Ricker:2015} is designed to conduct an all-sky transit survey of bright, nearby stars, providing an ideal data set for detecting short-period transiting super Earths. Owing to its high photometric precision and continuous monitoring over $\sim$27-day sectors, TESS is particularly sensitive to planets with orbital periods of days to weeks, a regime that is challenging for RV surveys alone. Notably, the first exoplanet discovered by TESS, $\pi$ Mensae c \citep{Huang:2018, Gandolfi:2018}, is a short-period super Earth orbiting a Sun-like star that also hosts a known long-period cold Jupiter \citep[HD 39091~b,][]{Jones:2002}. Another similar system was also found later by \citet{Teske:2020}. In both cases, the inner super Earths were not detected by RV alone prior to the TESS transit, even though intensive RV measurements had been conducted.

Motivated by these archetypal systems, we attempt to investigate the occurrence of transiting super Earths in systems with known RV cold Jupiters using TESS photometry. Compared to previous studies based on either transit samples or RV surveys alone, our approach combines the strengths of both techniques: the well-characterized RV sample provides a clean census of outer giant planets, while TESS enables a uniform and sensitive search for short-period SEs around the same host stars. This strategy allows us to directly constrain the probability of having inner super-Earths, especially the low-mass ones, in cold-Jupiter systems, while explicitly accounting for transit and detection completeness, thereby offering a more robust observational test of the SE-CJ correlation.

This paper is organized as follows: In Section \ref{sec:sample}, we describe our sample of systems with RV CJ; in Section \ref{sec:analysis}, we describe our pipeline to search for transiting SEs in TESS data and quantify the detection completeness; in Section~\ref{sec:validation}, we confirm the planetary nature of two new transiting SE candidates around HD~50554; in Section \ref{sec:result}, we estimate the occurrence rate of SE in our sample; in Section \ref{sec:summary}, we summarize our findings; in Section~\ref{sec:discussion}, we discuss the implications of our results to the super Earth--cold Jupiter connection.

\section{The RV Cold Jupiter sample} \label{sec:sample}

The sample used in this work is drawn from the NASA Exoplanet Archive (NEA, \citealt{Christiansen:2025}; retrieved on Nov.\ 16, 2025), {and we adopt the parameters with \texttt{default} label}. We select cold Jupiters satisfying the following criteria:
\begin{enumerate}[topsep=1pt, parsep=1pt]
    \item The host has no known stellar companions, as indicated by the NEA label \texttt{sy\_snum}. This is to remove binaries or multi-star systems {because} they may complicate the transit search and analysis, {as well as the interpretation of any observed SE--CJ correlation in the context of planetary formation and dynamical evolution.}
    \item The host is a Sun-like main-sequence star, defined to have a stellar mass in the range $0.6\,M_\odot < M_\star < 1.4\,M_\odot$ and a stellar radius below $2\,R_\odot$. For a few entries that do not have values for stellar mass or radius, we manually checked their status based on the discovery papers. Focusing on Sun-like main-sequence hosts allow us to have direct comparisons with previous studies.
    \item The discovery method for the planet is ``Radial Velocity''.
    \item The planet is qualified as a cold Jupiter, defined to have a minimum mass in the range $0.3\,M_{\rm J} < M_{\rm p}\sin i < 20\,M_{\rm J}$, \footnote{{We adopt the \texttt{pl\_bmassj} parameter, which represents the true mass if available, or the minimum mass ($M\sin i$) otherwise.}} orbital period $P > 300\,{\rm d}$, and semi-major axis $a < 10\,{\rm au}$.
    \item The planet is not discovered as part of follow-up observations of transiting systems. This is achieved by excluding systems with any of the following keywords in the name: WASP, HAT, Kepler, TOI, TIC, K2, KELT, and CoRoT.
    \item The planet has no hot ($P < 10\,{\rm days}$) or warm ($10\,{\rm d} < P < 300\,{\rm d}$) Jupiter companions ($m\sin i>0.3\,M_{\rm J}$). Similar to the previous criterion, the cold Jupiters beyond HJ or WJ orbits may have been detected through targeted follow-up rather than blind searches. {In addition, the presence of HJs and WJs may introduce additional correlations with inner small planets, which could complicate the interpretation of the SE--CJ relation and bias our inference.}
    \item The host star has TESS SPOC light curves {\citep{Jenkis:2016}}, and there are no bright ($\Delta m_G<5$\,mag) contaminants within the TESS aperture according to Gaia DR3 \citep{Gaia_cal:2023}. The existence of such nearby bright stars would complicate the validation of any potential transit signals.
\end{enumerate}
Our sample does not include several systems known to host both transiting SEs and CJs, most notably 55 Cancri and HD 191939. The 55 Cancri system has a stellar companion and hosts a warm Jupiter with an orbital period of $14$~days \citep{Butler:1997}. HD 191939 is excluded because its cold Jupiter was discovered only after the discovery of the inner transiting planets, and it also hosts a warm Jupiter with an orbital period of $102$~days \citep{Badenas:2020, Lubin:2022}.

\begin{figure*}
    \centering
    \includegraphics[width=\linewidth]{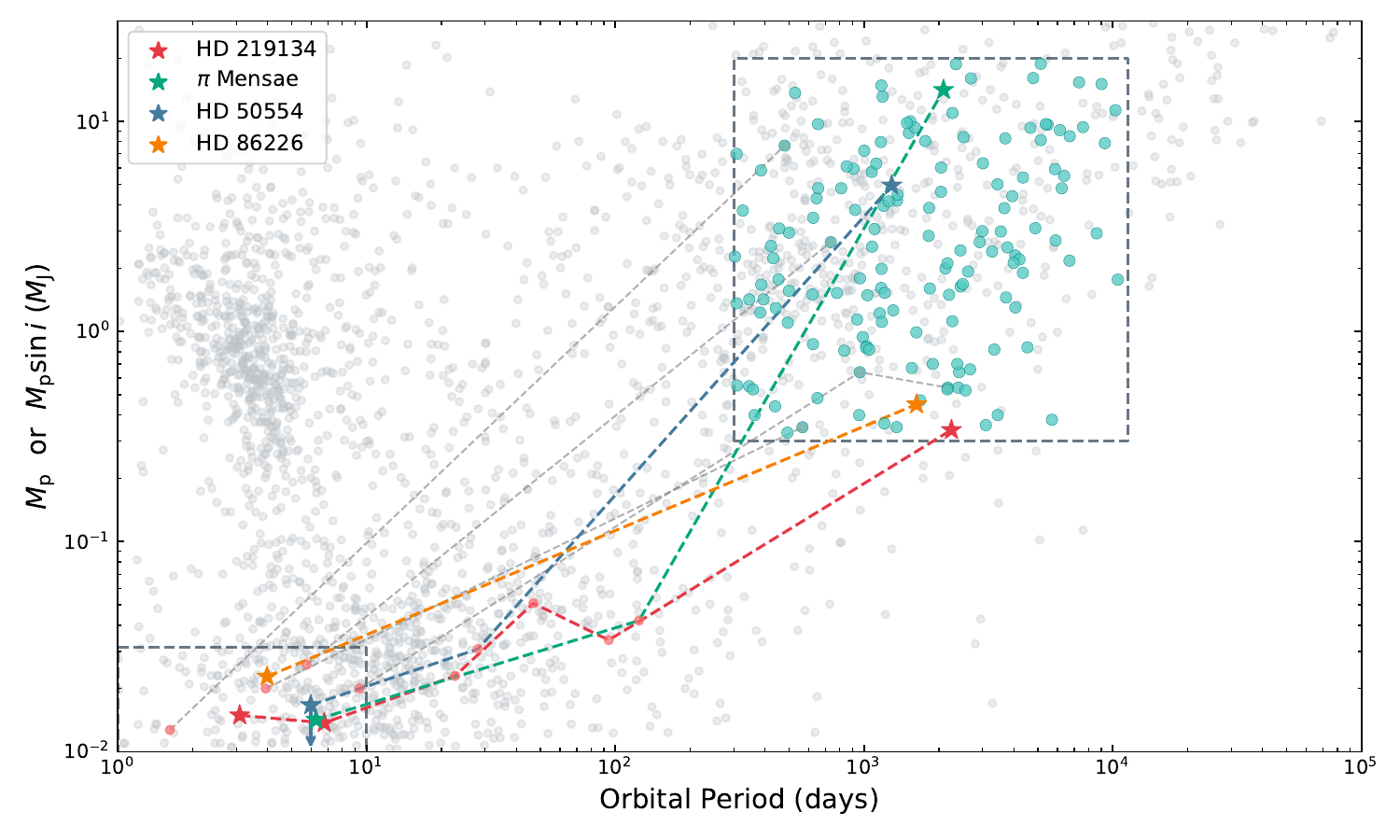}
    \caption{Planet mass (or minimum mass) as a function of orbital period for planetary systems hosting cold Jupiters. Gray points represent the full population of confirmed exoplanets from the NASA Exoplanet Archive. The dashed box outlines the parameter space adopted for CJs and HSEs in this work. Colored stars denote systems in our sample that host both a CJ and at least one inner SE detected by TESS: HD 219134, $\pi$ Mensae, HD 50554, and HD 86226. Cyan circles indicate CJs in our RV-selected sample without detected transiting inner SEs. Other HSEs detected via RV, along with planets in their systems, are marked by pink circles. Planets belonging to the same system are connected by dashed lines.}
    \label{fig:sample}
\end{figure*}

\begin{figure*}
    \centering
    \includegraphics[width=\linewidth]{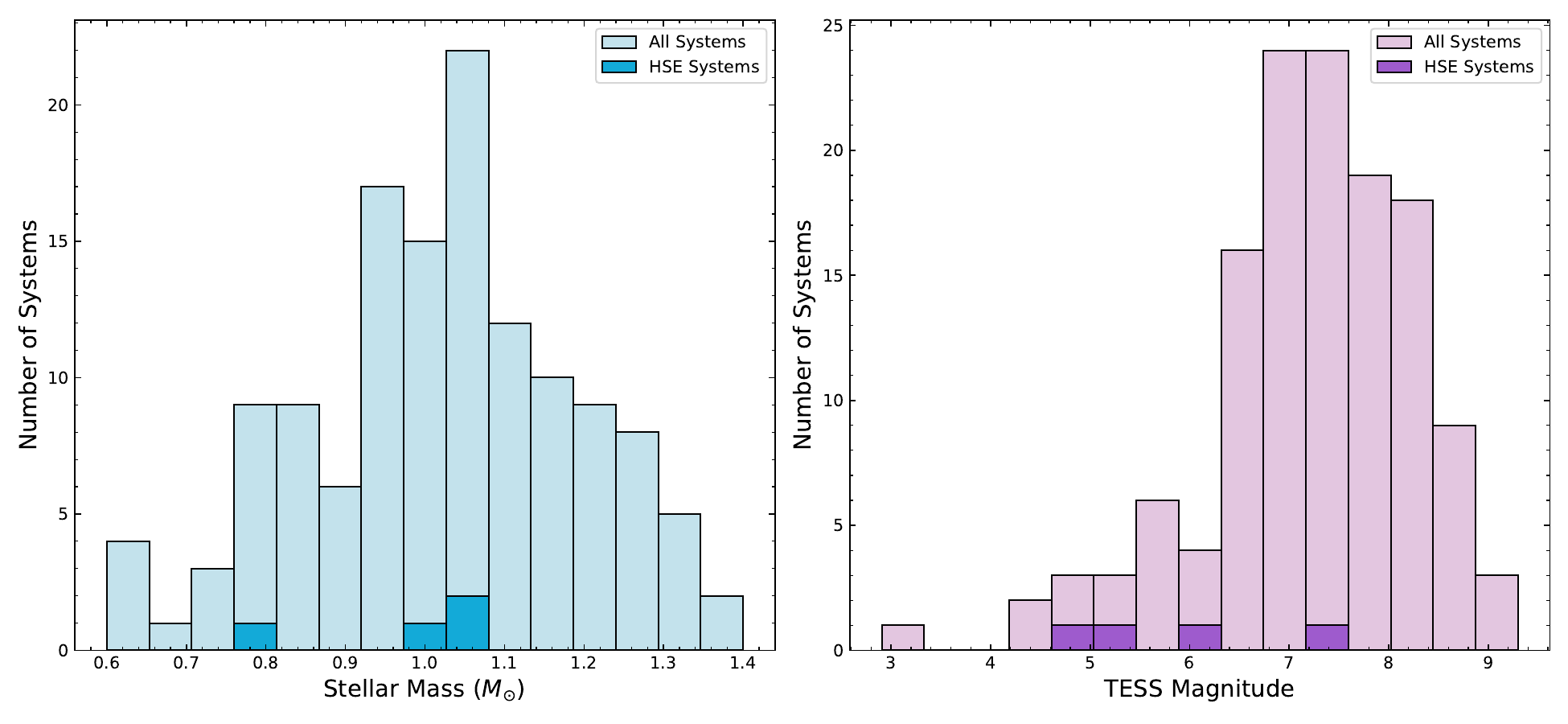}
    \caption{The distribution of the cold Jupiter host stars in our sample. {The host stars with transiting HSEs are marked in relatively dark color.} Left: The histogram of stellar mass. Right: The histogram of TESS magnitude.}
    \label{fig:mass_mag}
\end{figure*}

Finally, our sample contains 132 systems hosting 149 CJs. The host name is provided in Table \ref{tab:sample}, and these systems are shown in Figure~\ref{fig:sample}. Distributions of the stellar masses and TESS magnitudes of these systems are shown in Figure~\ref{fig:mass_mag}. The majority of the stars are truly solar-like with masses between 0.8--1.2$\,M_\odot$. These stars are relatively bright, with the overwhelming majority having TESS magnitude $<8$, whereas, for comparisons, TOIs found from the TESS primary mission typically have magnitudes in the range 8--10 \citep{Guerrero:2021}. This bias towards bright stars is understandable, as almost all of the stars in our sample are targets of RV surveys, which preferentially select bright Sun-like stars \citep{Wright:2012}.

A cross-match with the TESS Object of Interest \citep[TOI,][]{Guerrero:2021} table shows that five out of 132 systems contain transiting super Earths, including three with confirmed planets: HD 39091 ($\pi$ Mensae, \citealt{Huang:2018, Gandolfi:2018}), HD 219134 \citep{Gillon:2017, Motalebi:2015},
\footnote{Unlike the HD 191939 system, the cold Jupiter in HD 219134 was first detected, and the transiting signal of the inner small planet was detected later \citep{Vogt:2015, Motalebi:2015, Gillon:2017}.}
and HD 86226 \citep{Teske:2020}. There are additionally two systems hosting TESS planet candidates: HD 50554 (TOI-6965.01 and 6965.02) and HIP 54597 (TOI-6709.01). Almost all of the TESS transiting planets in these systems fall within the hot super-Earth (HSE for short hereafter) regime, with orbital periods shorter than 10 days and radii between 1--4$\,R_\oplus$, except for one candidate, TOI-6965.02 in HD 50554. In the next section, we will carry out our own transit search and quantify its detection efficiency.

\begin{deluxetable}{lccl}
\tablecaption{Cold Jupiters catalog of the sample.}
\label{tab:sample}
\tablehead{
\colhead{System name} & \colhead{TIC} & \colhead{$N_{\rm pl}$} & \colhead{TESS sectors}}
\startdata
HD 100777 & 49427341 & 1 & 6, 7, 87 \\
HD 102843 & 79290210 & 1 & 4, 31\\
HD 10442 & 380908091 & 1 & 7, 34\\
HD 105779 & 347713631 & 1 & 12 \\
HD 106252 & 397558558 & 1 & 10, 37 \\
HD 10647 & 229137615 & 1 & 29 \\
$\cdots$ & $\cdots$ & $\cdots$ & $\cdots$ \\
\enddata
\tablecomments{This table is published in its entirety in the machine-readable format. Only a portion of it is shown here for guidance regarding its form and content.}
\tablecomments{Here $N_{\rm pl}$ denotes the number of known planets in the system, as reported in NEA.}
\end{deluxetable}

\section{TESS Transit Analysis}\label{sec:analysis}

\subsection{Transit search}\label{sec:pipeline}
To guarantee the consistency of the statistical analysis and the completeness correction, we build our own pipeline to search for the SE transit signals in TESS data. For each target in our sample, we first retrieve the TESS Presearch Data Conditioning Simple Aperture Photometry (PDCSAP) light curve, processed by the Science Processing Operations Center (SPOC; \citealt{Jenkis:2016}) pipeline, using the \texttt{lightkurve} package \citep{lightkurve:2018}. The data are accessed as of November 25, 2025. {We detrend the light curves using the \texttt{biweight} method implemented in \texttt{wotan} \citep{Hippke:2019}, adopting a window length of $0.5\,\mathrm{d}$, which is more than three times longer than the typical transit duration to avoid overfitting and suppressing potential transit signals.}
We also clip outliers that are more than three times the median absolute deviation (MAD) from the median value within the same window length. 

Then we conduct the box least square (BLS; \citealt{Kovacs:2002}) search in order to identify the transit signals. The BLS search is performed on a two-dimensional grid, with 10,000 and 100 equally spaced values along orbital period between 1-10 days and transit duration between 0.01-0.3 days, respectively. We adopt the threshold-crossing event (TCE) criteria used in the TESS pipeline \citep{Guerrero:2021}, which define that a periodic signal is considered detected if it satisfies both (1) a BLS peak significance $>9$, and (2) a signal-to-pink-noise ratio S/N$_{\rm pink}>9$. The BLS peak significance is computed as the height of the maximum BLS peak relative to the noise floor of the periodogram, defined as the median BLS power within a $\pm0.3$-day window around the peak. The period and transit duration values corresponding to the highest BLS significance are used to calculate S/N$_{\rm pink}$, which is defined as
\begin{equation} \label{eqn:snr_pink}
    \mathrm{S} / \mathrm{N}_{\mathrm{pink}}=\frac{\delta}{\sqrt{\left(\sigma_w^2 / n_{\rm d}\right)+\left(\sigma_r^2 / N_{\rm tr}\right)}} .
\end{equation}
Here $\delta$ is the transit depth, and $\sigma_w$ and $\sigma_r$ are the white and red noises of the light curve, respectively. Quantities $n_{\rm d}$ and $N_{\rm tr}$ are the number of in-transit points and the number of transits, respectively \citep{Pont:2006, Hartman:2016}. We follow the procedure of \cite{Kunimoto:2025} in computing each of those parameters in Equation~(\ref{eqn:snr_pink}).

Out of the 132 systems with RV cold Jupiters, our pipeline detects five transit candidates, as listed in Table~\ref{tab:HSE}. Four of them correspond to TESS-confirmed planets: $\pi$ Men c, HD 86226 c, and HD 219134 b and c. The fifth one is the TESS candidate TOI-6965.01 around HD 50554. Two other TESS transit candidates that were mentioned in Section~\ref{sec:sample}, namely TOI-6965.02 around HD 50554 and TOI-6709.01 around HIP 54597, are not detected by our pipeline. For TOI-6965.02, its orbital period of 28 days exceeds the period range explored by our pipeline, and TOI-6709.01 has a low transit S/N that did not pass our detection criteria. Since these two candidates are not detected by our pipeline, we do not include them in our transiting SE sample. Given the unconfirmed status of TOI-6965.01, we will first validate its planetary nature in Section~\ref{sec:validation} before proceeding to the sample statistics. The second candidate in the same system, TOI-6965.02, is also validated, even though it will not be included in the statistical analysis. 

Besides the transit signals, our pipeline also identifies three systems—--HD 165131, HD 70642, and HD 80883--—whose light curves exhibit periodic signals that formally satisfy our detection criteria. However, inspection of the phase-folded light curves reveals sinusoidal variations rather than transit-like features, indicating that these signals are not of planetary transit origin. Such sinusoidal trends were not removed by our detrending procedure because their periods ($\sim0.2$~days) are shorter than the 0.3-day window used for detrending. We can in principle remove these three systems from our sample because of the false signals from the sinusoidal variations. However, these targets have otherwise low photometric noise levels, and our injection--recovery tests show that the injected transit signals are unaffected by the sinusoidal photometric behavior and thus detectable nevertheless. Given this and the fact that these systems are rare among our sample, we therefore retain these three targets in our final sample.

\begin{deluxetable}{lcccc}
\tablecaption{TESS transiting super Earths in the RV cold Jupiter systems of our sample.}
\label{tab:HSE}
\tablehead{
\colhead{Planet name} & \colhead{TOI} & \colhead{Period (days)} & \colhead{$R_p\,(R_\oplus)$} & \colhead{Ref.}}
\startdata
\makecell{HD 39091 c\\($\pi$ Men c)} & 144.01 & $6.26790\,(46)$ & $2.04\,(5)$ & 1 \\
HD 86226 c & 652.01 & $3.98442\,(18)$ & $2.16\,(8)$ & 2 \\
HD 219134 b & 1469.01 & $3.0935\,(3)$ & $1.61\,(9)$ & 3 \\
HD 219134 c & 1469.02 & $6.76458\,(33)$ & $1.511\,(47)$ & 4 \\
HD 50554 c & 6965.01 & $5.969362\,(20)$ & $1.31\,(6)$ & 5  \\
\enddata
\tablecomments{References: (1) \citet{Huang:2018}, (2) \citet{Teske:2020}, (3) \citet{Motalebi:2015}, (4) \citet{Gillon:2017}, (5) This work}
\end{deluxetable}

\subsection{Detection efficiency calculation}

\begin{figure*}
    \centering
    \includegraphics[width=\linewidth]{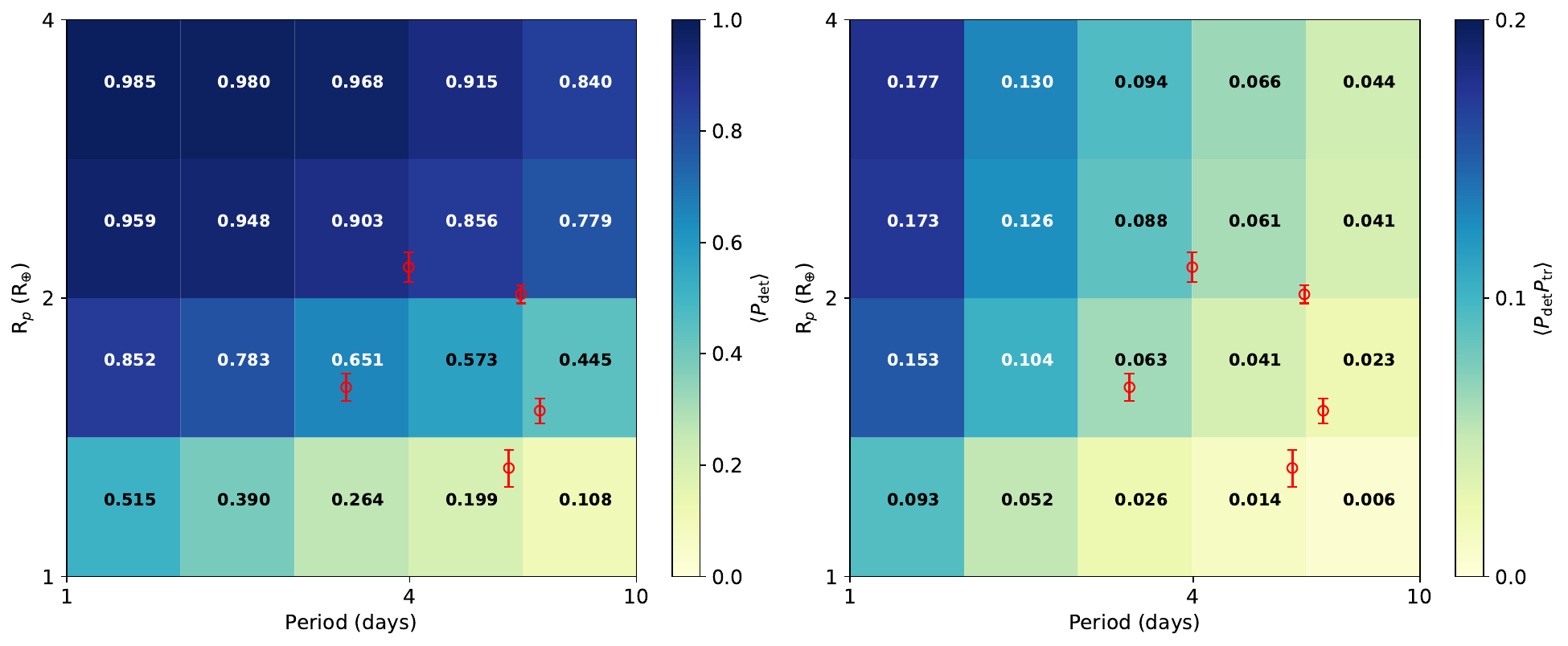}
    \caption{The average detection sensitivity map (left panel) and the full completeness (i.e., detection sensitivity \& transit probability) map (right panel) as a function of orbital period and planet radius. The number in each cell represents the average detection probability across all 132 targets. The red circles with error bars indicate the five HSEs detected by our pipeline.}
    \label{fig:sen_map}
\end{figure*}

We quantify the detection efficiency of our search pipeline through an injection–-recovery experiment. For systems with known transiting planets, we remove their transit signals based on the best-fit parameters from NEA, and we manually check this step and make sure that no peak in the BLS periodogram has significance above nine.

For each target, we inject 200 synthetic transit signals into the raw light curve. Similar to \citet{Gan:2023}, planetary radii are randomly drawn from 1--4 $R_\oplus$, orbital periods are sampled from a log-uniform distribution between 1--10 days, and impact parameters are randomly drawn with $b \leq 0.9$. The epoch of the first transit is randomly assigned between the start of the observations, $t_{\rm begin}$, and $t_{\rm begin}+P$. For simplicity, we adopt fixed quadratic limb-darkening coefficients $(u_1, u_2) = (0.3, 0.3)$ for all injections. Synthetic transit models are generated using the analytic formalism of \citet{Mandel:2002}, as implemented in the \texttt{batman} package \citep{batman:2015}. We correct for the dilution effect by modifying the injected radius ratio $R_p/R_\star$ to $(R_p/R_\star)/\sqrt{(1+A_D)}$ \citep{Gan:2023, Espinoza:2019}, where $A_D$ is the contamination ratio reported by TIC v8 \citep{Stassun:2019}. These models are then injected into the raw light curves and passed through our full transit-search pipeline to evaluate the recovery efficiency.

During the recovery stage, we largely follow the search procedure described in Section~\ref{sec:pipeline}, including the detrending and detection process, except for one minor revision. In estimating S/N$_{\rm pink}$, we adopt directly the value of the injected orbital period rather than performing the grid search. This obviously speeds up the process, and it also avoids the occasional situations where the most significant BLS peak may be mistakenly attributed to harmonics of the injected period or the sinusoidal variations (for the three systems discussed in Section~\ref{sec:pipeline}). 

The injection–recovery experiment is done for all target stars in our sample, and the detection sensitivity is defined as the ratio between recovered signals and all injected signals. The sensitivity map is derived on a $5\times4$ grid in the orbital period--radius plane, as is shown in the left panel of Figure~\ref{fig:sen_map}. No validations of the injected signals were performed in our experiment, as our sample is quite clean and exempt from any severe false positives. The only exception is the three targets with sinusoidal variations, but their effect has been mitigated in the signal evaluation step. 
Figure~\ref{fig:sen_map} shows that TESS achieves a typical detection sensitivity of roughly $\sim50\%$ for planets with radii between 1--4$\,R_\oplus$ and orbital periods between 1--10\,days around the RV targets with cold Jupiters. 

The geometric transit probability is evaluated as
\begin{equation} \label{eqn:p_tr}
    p_{\mathrm{tr}}=0.9 \frac{R_*}{a}=0.9 R_*\left(\frac{G M_* P^2}{4 \pi^2}\right)^{-1 / 3}.
\end{equation}
The factor of 0.9 comes from considering only transit planets with $b<0.9$ in this study. Taking the transit probability into account, the search completeness map is shown in the right panel of Figure~\ref{fig:sen_map}. The above evaluation has assumed an isotropic distribution for the orbital inclination of the transiting super Earths, which we argue remains a reasonable assumption even though these systems have RV cold Jupiter detections. On one hand, the orbital inclinations of the RV cold Jupiters may deviate from an ideal isotropic distribution because of the RV detection bias. On the other hand, the mutual inclinations between the inner small planets and the outer giants, although small on average \citep{Masuda:2020}, depend on the planet multiplicity and can be significant for systems with only one or two planets \citep{Herman:2019}. Detailed analyses of systems like $\pi$ Mensae indeed confirm the presence of large mutual inclinations \citep{Xuan:2020, DeRosa:2020, Damasso:2020}. More discussions on the latter point will be given in Section~\ref{sec:cj-mass}.

\section{Validation of two Super Earths around HD 50554}\label{sec:validation}

\subsection{TESS observations \& light curve modeling}

\begin{figure}
    \centering
    \includegraphics[width=\linewidth]{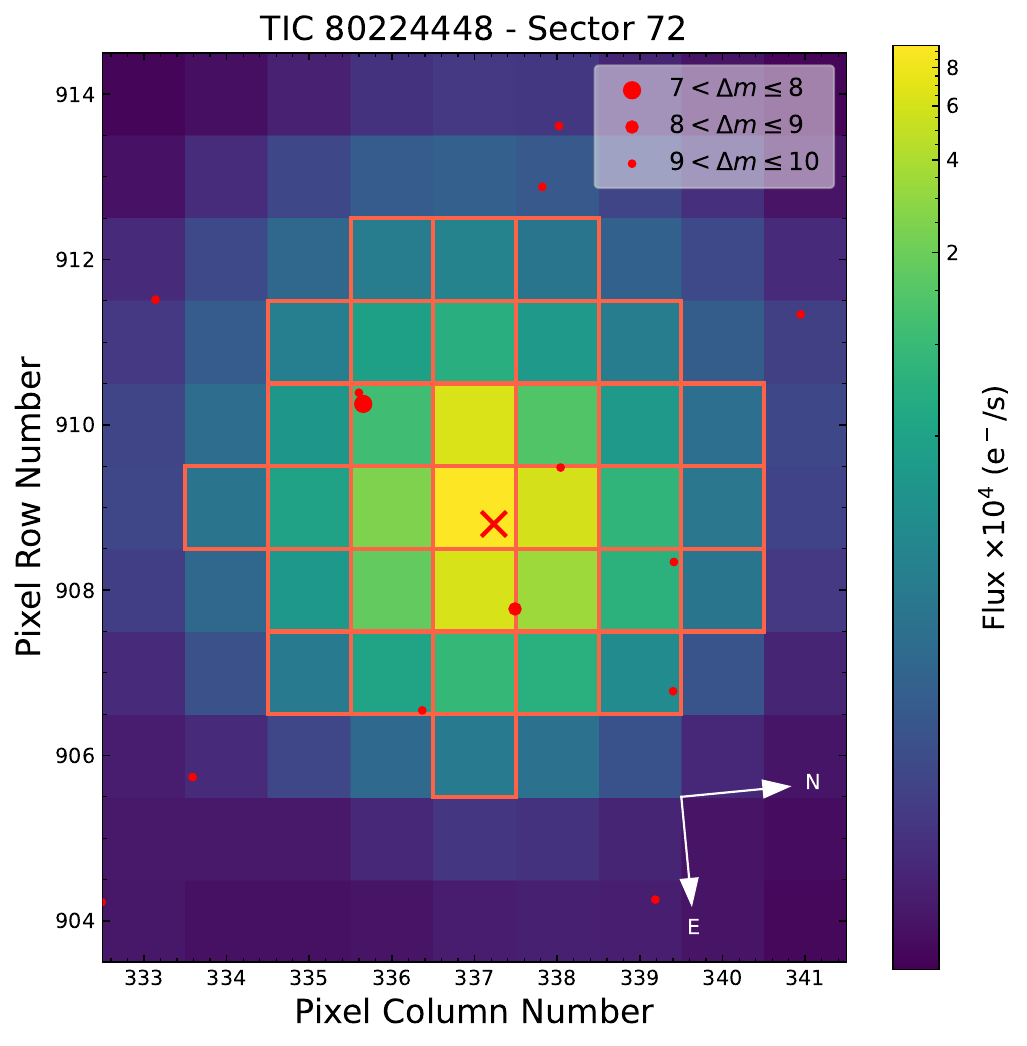}
    \caption{TESS TPF image of HD 50554 (TIC 80224448) observed in Sector 72. Orange squares outline the photometric apertures. The red cross marks the nominal position of the target star. Red dots denote nearby Gaia sources, and the sizes of dots reflect their magnitude difference compared to the HD 50554.}
    \label{fig:tpf}
\end{figure}

\begin{figure*}
    \centering
    \includegraphics[width=\linewidth]{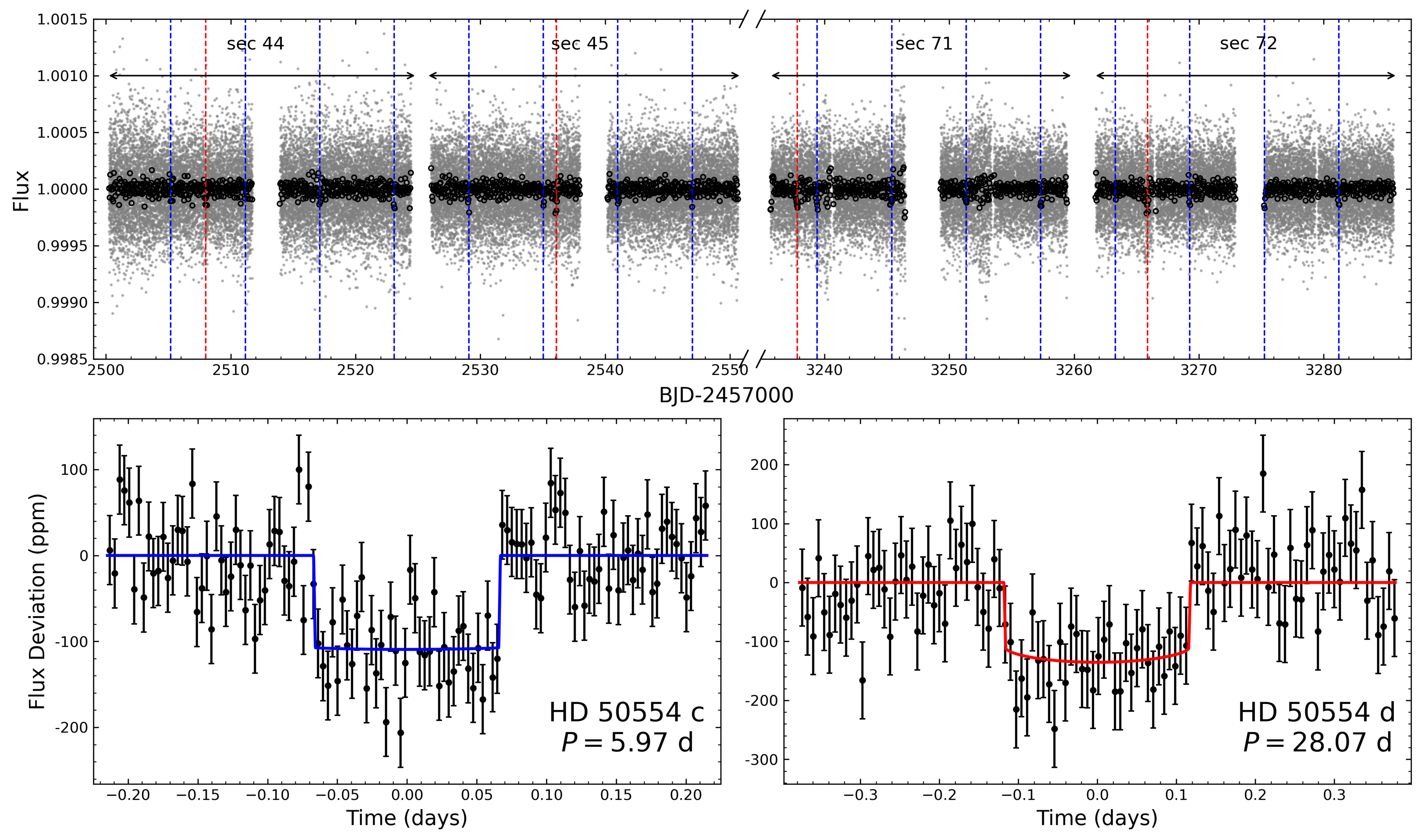}
    \caption{The TESS light curves of HD 50554. The top panel shows the flattened TESS light curves from Sectors 44, 45, 71, and 72. Grey points indicate short-cadence data with 2-minute sampling, and black points are 1-hour averages for visualization. Blue and red dashed lines mark the predicted mid-transit times of HD~50554~c (TOI-6965.01) and d (TOI-6965.02), respectively. The bottom panels show the phase-folded light curve, along with the best-fitting model of HD 50554 c (left) and HD 50554 d (right). The black dots represent 5-minute averages in the left and 10-minute averages in the right.}
    \label{fig:lc}
\end{figure*}

HD 50554 is a Sun-like star with TESS magnitude of $6.31$ (see Table~\ref{tab:pars} for more details).
TESS observed this star at a 20-second cadence in Sectors 44, 45, 71, and 72. The target pixel file (TPF) and photometric aperture used in Sector 72 are shown in Figure \ref{fig:tpf}, which is made via \texttt{tpfplotter} \citep{tpfplotter:2020}. {This target was previously observed during Campaign 0 of the K2 mission \citep{Howell:2014}. However, with a Kepler magnitude of $6.8$, the star was severely saturated, and we were unable to recover any transit signals from the archival K2 data. So we only focus on the TESS data in the following analysis.}
According to Gaia DR3, the second brightest star in the aperture is still about 8 mag fainter than the target, {suggesting very low-level contamination. This is confirmed by the flux contamination level of only $0.2\%$ given by \texttt{TESS-cont} \citep{TESS-cont:2024}}. So we can ignore the light contamination from nearby stars. Data acquisition and pre-processing are performed following the procedures given in Section \ref{sec:pipeline}. We confirmed the 5.97 d signal with a BLS significance of 10.8 and a S/N$_{\rm pink}$ of 13.4, both above the thresholds to be claimed as a detection. However, our pipeline is not suitable for searching for the second signal with 28.07 d period, as the detrending with a 0.3-day window may suppress the corresponding transit signal. To detrend the light curve while preserving the transit signals, we first mask out the transit events based on the NEA parameter values and then fit a Gaussian process (GP) model with the Matern-3/2 kernel using the \texttt{celerite} package \citep{celerite:2017}. The flattened light curve is shown in the top panel of Figure~\ref{fig:lc}.


We model the TESS light curve using the \cite{Mandel:2002} model as implemented in \texttt{Batman} \citep{batman:2015}. Over nearly 800 days covered by the four sectors, there are 16 transits of the inner candidate (TOI-6965.01) and four transits of the outer candidate (TOI-6965.02). We select only data with a 1-day length centered on the expected transit mid-point for light curve modeling. A quadratic limb-darkening law is adopted, and the re-parameterization into $q_1$ and $q_2$ according to \citet{Kipping:2013} is used. We use the Markov chain Monte Carlo (MCMC) method as implemented in \texttt{emcee} of \cite{Foreman-Mackey:2013} to sample the posterior distributions of the model parameters. The phase-folded light curves of two candidates are shown in the lower panels of Figure~\ref{fig:lc}). The median values and 1-$\sigma$ uncertainties of the model parameters determined from the posterior distributions are summarized in Table~\ref{tab:pars}. 

According to our model, the two transiting planet candidates around HD 50554 have radii of $1.31\pm 0.06\,R_\oplus$ and $1.41^{+0.14}_{-0.09}\,R_\oplus$ and orbital periods of $\sim6$ and $\sim28$ days, respectively. Therefore, they both fall below the radius valley and are thus likely rocky with no thick envelopes \citep{Fulton:2017, Owen:2017}.

\subsection{Mass constraints from archival RV}

The super Jupiter HD~50554~b, with minimum mass $\sim4\,M_{\rm J}$ and semi-major axis $\sim2.3\,$au around HD 50554 was discovered independently by \citet{Fischer:2002} and \citet{Perrier:2003}. Since then, the star has been continuously observed for more than two decades, but no other RV signals have been detected \citep[e.g.,][]{Xiao:2023}. Given the estimated masses of $\sim2\,M_\oplus$ and $\sim3\,M_\oplus$ from the \citet{Chen:2017} mass--radius relation, the two transiting planets produce RV semi-amplitudes of 0.7 and 0.6 m/s, respectively. It is therefore not a surprise that these planets have not been detected.

\begin{figure}
    \centering
    \includegraphics[width=\linewidth]{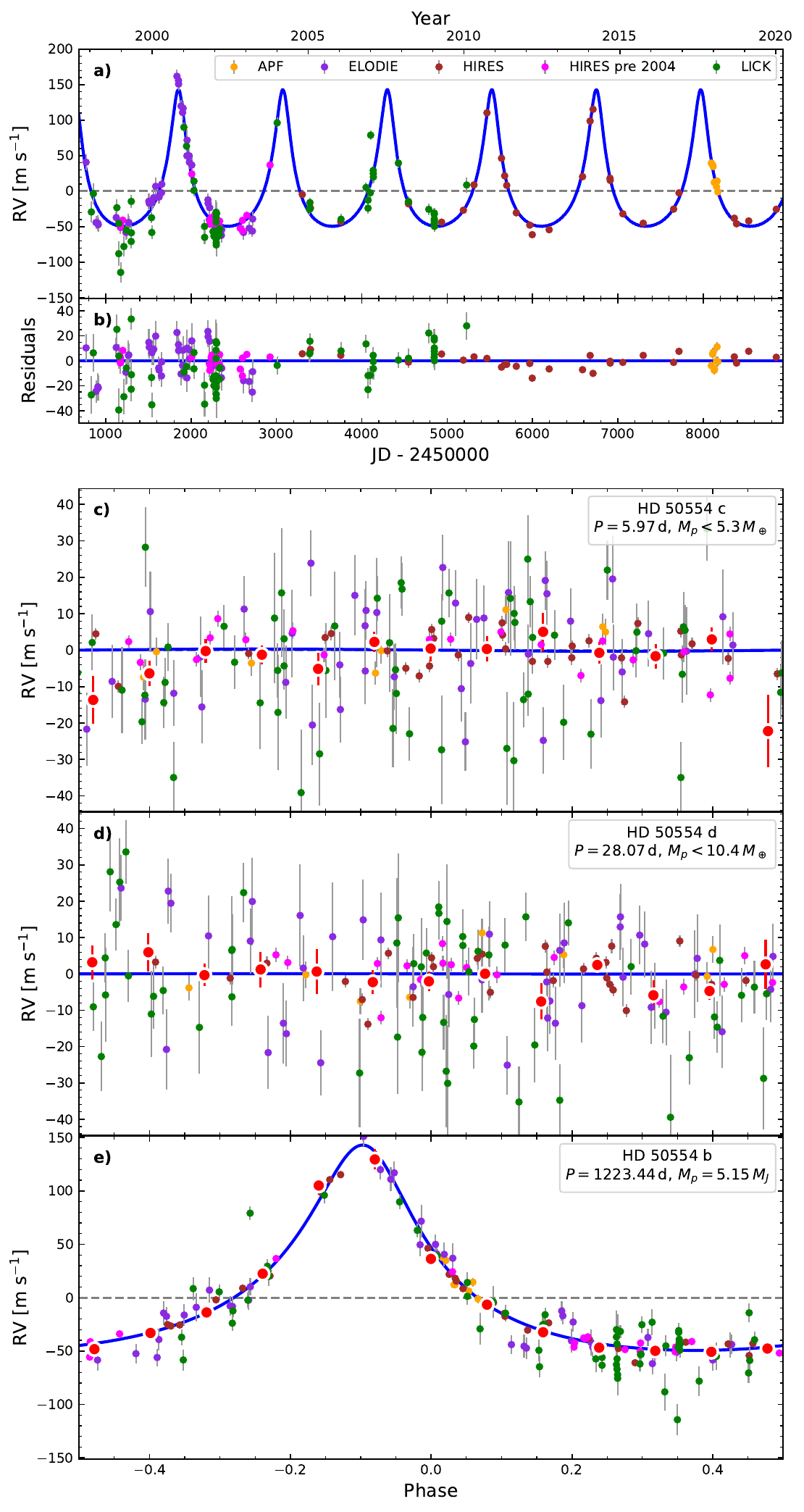}
    \caption{Radial-velocity analysis of the HD 50554 system plotted by \texttt{radvel}. (a-b) Current public RV time-series and residuals of the full three-planet model. (c–e) Phase-folded RV signals for planets c, d, and b, respectively. The blue curves show the corresponding best-fit orbital models.}
    \label{fig:rvplot}
\end{figure}

We nevertheless attempt to place limits on the masses of the two planet candidates. In this analysis, we include RV measurements from the California Legacy Survey (CLS, \citealt{Rosenthal:2021}), which collected a total of 157 observations from Keck/HIRES, APF/Levy, and Lick/Hamilton, and archival measurements from ELODIE \citep{Perrier:2003}, as presented in Figure \ref{fig:rvplot}. The RV modeling is done through the \texttt{radvel} package \citep{radvel:2018}. We first identify the model parameters corresponding to HD~50554~b and then subtract its signal from the RV data. We then fit the residuals to constrain the semi-amplitudes of the two inner candidates, assuming circular orbits for both planets. In this step, we have imposed Gaussian priors on the orbital periods and mid-transit epochs from the transit fit. These RV data yield 95\% upper limits of $5.3$ and $10.4\,M_\oplus$ for the 6-day and 28-day planets, respectively.

\begin{deluxetable}{lcc}
\label{tab:pars}
\tablecaption{Stellar and planetary parameters of the HD 50554 system}
\tablehead{
\colhead{\makecell{Stellar\\Parameters}} & \colhead{Value} & \colhead{Reference}}
\startdata
TIC ID               & 80224448 & 1 \\
TOI ID               & 6965 & \\
TESS mag        & 6.31 & 1 \\
$M_\star\,(M_\odot)$ & $1.04\, (5)$ & 2 \\
$R_\star\,(R_\odot)$ & $1.146 \, (26)$ & 2 \\
$\mathrm{[Fe/H]}$    & $-0.02\, (6)$ & 2 \\
$T_{\rm eff}$ (K)    & $5968 \, (96)$ & 2 \\
Distance (pc) & $31.070\, (23)$ & 3 \\
\tableline
\makecell{Planet\\Parameters} & Planet c & Planet d \\
\tableline
$T_c$ (TBJD) & $2505.1917\, (20)$ & $2508.005\,(9)$ \\
$P$ (days) & $5.969362\, (20)$ & $28.06940\, (40)$ \\
$R_p/R_\star$ & $0.01045\, (41)$ & $0.0113^{+0.0011}_{-0.0007}$ \\
$R_p$ ($R_\oplus$) & $1.31\, (6)$ & $1.41^{+0.14}_{-0.09}$ \\
$b$ & $0.49^{+0.29}_{-0.33}$ & $0.50\, (35)$ \\
$a/R_\star$ & $\mathbf{12.24\,(39)}$ & $\mathbf{34.4\,(1.1)}$ \\
$a$ (au) & $\mathbf{0.0652\,(10)}$ & $\mathbf{0.1831\,(29)}$ \\
$T_{14, \rm tr}\,({\rm hr})$\tablenotemark{a} &
$3.30^{+0.14}_{-0.09}$ & $6.21^{+0.64}_{-0.44}$ \\
$T_{12, \rm tr}\,({\rm hr})$\tablenotemark{a} &
$0.12^{+0.13}_{-0.08}$ & $0.73^{+0.59}_{-0.43}$ \\
$M_p$ ($M_\oplus$) & $<5.3$\tablenotemark{b} & $<10.4$\tablenotemark{b} \\
\enddata
\tablecomments{References: (1) \citet{Stassun:2019}, (2) \citet{Rosenthal:2021}, (3) \citet{Gaia_cal:2023}}
\tablenotetext{a}{From trapezoidal transit models.}
\tablenotetext{b}{The 95\% upper limit of planetary masses.}
\end{deluxetable}

\subsection{Transit signal validation}
We have searched for odd--even variations and searched for secondary eclipses in the TESS light curve, and both candidates pass these vettings. Together with the non-detection of the RV signals at the known orbital periods, these add further confidence that the transit signals are indeed produced by low-mass planets around the target star. Other false positive scenarios are also ruled out:
\begin{itemize}
\item HD 50554 has no stellar companion(s) that could explain the transit signals. HD 50554~b has a measured true mass of $\sim6\,M_{\rm J}$, based on a joint modeling to the Gaia and Hipparcos astrometric measurements \citep{Xiao:2023}. Apart from HD 50554~b, there is no other periodic signal in the RV data, and the lack of a linear trend in the RV residuals rules out any massive objects within $\sim30\,$au, which, combined with the known distance of HD~50554, corresponds to an angular separation of $\sim$1". The presence of a stellar companion at wider separations is ruled out by Gaia DR3 \citep{Gaia_cal:2023}. 

\item The transit events cannot be explained by background stars unassociated with HD 50554. As shown in Figure~\ref{fig:tpf}, there are no stars of comparable brightness ($\Delta m<5$) within the TESS aperture. In order to explain the $100\,$ppm transit signals on the 6.3 mag target, a nearby star brighter than $\sim16$ mag is required. According to Gaia DR3, four stars pass this threshold within the TESS aperture, and the brightest of them have $m_{\rm TESS}=14$.
\footnote{{In principle, there could be other background stars that are close to HD 50554 and yet unresolved by Gaia. However, given the low surface density of qualified stars around HD 50554, the probability for this to occur is extremely small ($\sim 4\times 10^{-4}$). Indeed, high-resolution speckle imaging taken by `Alopeke speckle instrument on the Gemini South telescope on HD 50554, which reaches a contrast of $\Delta m \sim 7$ mag at $0.5''$, reveals no additional star \citep{Howell:2025}.}}
Here we have converted the Gaia $G$ magnitudes to TESS magnitudes using the relations given in \citet{Stassun:2019}. For such faint stars to account for the transit signals, the eclipsing events would need to be intrinsically deep, with an estimated eclipsing depth of 
\begin{equation}
\delta_{\rm EB}=\delta\left(1+10^{0.4\Delta m_{\rm TESS}} \right) .
\end{equation}
For the faint star with $m_{\rm TESS}=14$, the eclipsing depths without dilution are 14\% and 16\% for the two candidate signals, respectively. Such deep eclipses would yield V-shaped light curves with large ingress-to-total-duration ratios ($T_{12}/T_{14}\gtrsim0.28$ and $\gtrsim0.30$, respectively). Such scenarios are effectively ruled out by our transit constraints (see Table~\ref{tab:pars}).
\end{itemize}

\begin{table*}
\centering
\caption{Sample sizes and enhancement factor under different selection criteria. The first case is our fiducial criterion as listed in Section \ref{sec:sample}, where CJ is defined as $0.3 < M_{\rm p}\sin i < 20\,M_{\rm J}$, $P > 300\,\rm days$ and $a < 10\,\rm au$, WJ is defined as $M_{\rm p}\sin i > 0.3\,M_{\rm J}$, $P < 300\,\rm days$. The brackets in the subsequent columns show the definitions that are different from the fiducial criterion.}
\label{tab:definition}
\begin{tabular}{lcc}
\hline\hline
Selection criterion & Sample size & Enhancement factor \\
\hline
Fiducial criterion & 4/132 & $6.5^{+3.1}_{-2.3}$ \\ CJ definition ($0.3 < M_{\rm p}\sin i < 13\,M_{\rm J}$) & 3/124 & $5.7^{+3.0}_{-2.2}$ \\
CJ definition ($1 < a < 10\,\rm au$) & 4/127 & $6.7^{+3.2}_{-2.4}$ \\
CJ definition ($1 < a < 10\,{\rm au},\,0.3 < M_{\rm p}\sin i < 13\,M_{\rm J}$) & 3/118 & $6.0^{+3.2}_{-2.3}$ \\
WJ definition ($P < 100\,\rm days$) & 4/141 & $6.1^{+2.9}_{-2.2}$ \\
\hline
\end{tabular}
\end{table*}

\section{Super Earth occurrence rate enhanced by the presence of cold Jupiters}\label{sec:result}

Our sample contains 132 Sun-like stars hosting RV-detected cold Jupiters, among which four contain in total five transiting super Earths with orbital periods below 10\,d. These numbers strongly suggest that the occurrence of hot super Earths is enhanced by the presence of cold Jupiters, as is explained below.

In the absence of any correlation between HSEs and CJs, the expected number of transiting HSEs should follows their intrinsic occurrence rate, which can be estimated as
\begin{equation} \label{eqn:N_HSE}
    \bar{N}_{\rm HSE} = N_\star \sum_{i, j} \eta_{{\rm HSE}, ij} \cdot p_{{\rm tr}, ij} \cdot p_{{\rm det}, ij}.
\end{equation}
Here $N_\star=132$ is the total number of stars in the sample, $\eta_{\rm HSE}$ denotes the HSE occurrence rate in the period vs.\ radius plane, $p_{\rm tr}$ is the geometric transit probability (Equation~\ref{eqn:p_tr}), and $p_{\rm det}$ is the pipeline detection efficiency map (left panel in Figure~\ref{fig:sen_map}). The summation is performed over both orbital period (index $i$) and planetary radius (index $j$) directions. 

{We use the HSE intrinsic occurrence rates from \citet{Zhu:2021} and \citet{Petigura:2018}, which were derived based on the \textit{Kepler} sample. The HSE intrinsic occurrence rate exhibits a weak positive correlation with stellar metallicity, $\eta({\rm HSE})\propto 10^{\beta{\rm [Fe/H]}}$ with $\beta= 0.6\pm0.2$ \citep{Petigura:2018} While the \textit{Kepler} sample of solar-type stars has a near-solar average metallicity \citep{Dong:2014}, our RV cold Jupiter sample is systematically more metal-rich (see Figure~\ref{fig:metal}). This average metallicity enhancement of $\Delta {\rm [Fe/H]}\approx 0.2\,$dex consequently elevates the expected HSE occurrence rate in our sample by a factor of $1.3\pm 0.1$. Applying this metallicity correction to the baseline occurrence rate distributions from \citet{Zhu:2021} and \citet{Petigura:2018}, we obtain an expected HSE number of $\bar{N}_{\rm HSE}=0.87\pm0.07$ and $\bar{N}_{\rm HSE}=0.68\pm 0.06$ according to Equation~(\ref{eqn:N_HSE}), respectively. Here the uncertainty is dominated by the metallicity enhancement factor. 
Both predicted values are substantially smaller than the number of actual detections ($N_{\rm HSE}=5$).}

\begin{figure}
    \centering
    \includegraphics[width=\linewidth]{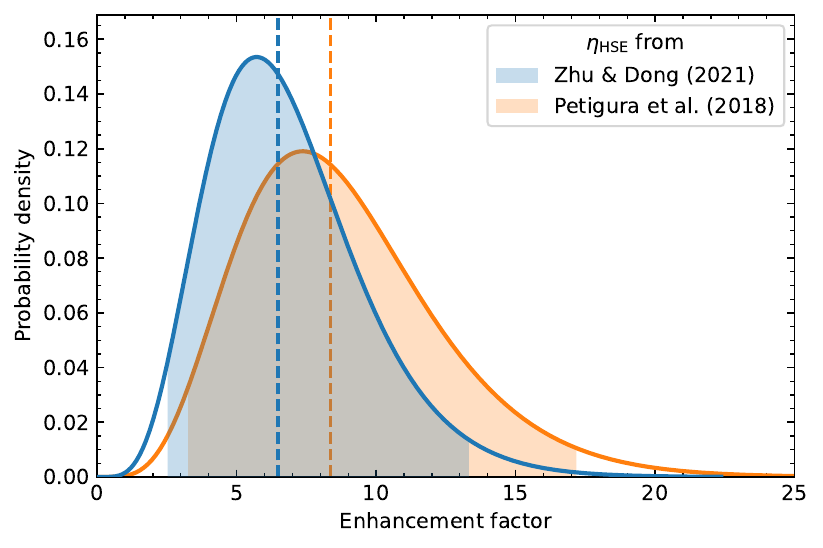}
    \caption{Posterior distributions of the enhancement factor of the occurrence rate of hot SE in the presence of CJ. The two distributions shown in different colors are derived using the occurrence rates of HSEs from \cite{Zhu:2021} and \cite{Petigura:2018}, respectively. The median values are indicated by the vertical dashed lines, and the 95\% confidence interval about the median is indicated by the shaded region.}
    \label{fig:enhancement}
\end{figure}

To quantify the impact of CJ on the occurrence rate of HSE, we define an enhancement factor $f$
\begin{equation} \label{eqn:enhance-factor}
    f \equiv \frac{\eta({\rm HSE|CJ})}{\eta({\rm HSE})} .
\end{equation}
With a Poisson likelihood assumed for the number of HSE detections and a flat prior adopted on $f$, the posterior distribution of the enhancement factor $f$ given $N_{\rm HSE}$ and $\bar{N}_{\rm HSE}$ follows a Gamma distribution with a shape parameter $N_{\rm HSE}+1$ and a rate parameter $\bar{N}_{\rm HSE}$ 
\begin{equation}
    P\left(f|N_{\rm HSE}, \bar{N}_{\rm HSE}\right) 
    = \frac{\bar{N}_{\rm HSE}^{\,N_{\rm HSE}+1}}{\Gamma(N_{\rm HSE}+1)} 
    f^{N_{\rm HSE}} e^{-f \bar{N}_{\rm HSE}} .
\end{equation}
{The uncertainty of $\bar{N}_{\rm HSE}$ is relatively small and can be ignored in deriving the distribution of enhancement factor}. With $\bar{N}_{\rm HSE}=0.87$ and $N_{\rm HSE}=5$, the enhancement factor is constrained to be $f = 6.5^{+3.1}_{-2.3}$ (68\% confidence interval). The posterior distribution of the enhancement is shown in Figure \ref{fig:enhancement}. The probability of $f\leq1$ is only 0.1\%, which implies we can safely rule out that the presence of CJs will suppress the existence of HSE at high statistical significance ($p<0.001$). This result is insensitive to the adopted HSE occurrence rate map. For example, adopting the $\bar{N}_{\rm HSE}=0.68$ derived from \citet{Petigura:2018}, we obtain an enhancement factor of $f = 8.4^{+4.0}_{-3.0}$. These consistent results suggest that the presence of CJs does enhance the occurrence of HSEs by a factor of a few.

The above enhancement factor is insensitive to our definitions of CJ and/or WJ. If we apply different definitions of CJ and WJ in our sample selection, the sample size is changed slightly, and the derived enhancement factors remain largely unchanged, as is shown in Table~\ref{tab:definition}.

With the known value of $\eta({\rm HSE})=0.176$ from \citet{Zhu:2021}, we constrain the occurrence rate of HSE around Sun-like stars with at least one cold Jupiter,
\begin{equation}
    \eta({\rm HSE|CJ}) = 1.13^{+0.55}_{-0.41},
\end{equation}
and $\eta({\rm HSE|CJ}) = 1.25^{+0.61}_{-0.46}$ using $\eta({\rm HSE})=0.150$ from \citealt{Petigura:2018}.

We can go a step further to derive the fraction of Sun-like stars with at least one HSE given that there is already at least one cold Jupiter, $P({\rm HSE|CJ})$. The distinction between this and $\eta({\rm HSE|CJ})$ is in the planet multiplicity. Super Earths are frequently found in multi-planet systems \citep{Lissauer:2011, Fabrycky:2014}, and each system may contain on average $\sim3$ such planets within 1\,au \citep{Zhu:2018b}. Although the average multiplicity of hot super Earths is not known, we can take the observed average multiplicity of our sample ($5/4=1.25$) as an appropriate estimate. Then we have 
\begin{equation}\label{eqn:con_prob}
    P({\rm HSE|CJ})\approx 91\%.
\end{equation}
That is, about 91\% of Sun-like stars with cold Jupiters also host hot super Earths.

\section{Summary} \label{sec:summary}

This work studies the correlation between cold Jupiters and hot super Earths. We uniformly selected 132 systems with Sun-like hosts and at least one known cold Jupiter detected via RV. To avoid the potential biases associated with hot and warm Jupiters, cold Jupiters with such inner friends were excluded from the sample. We then systematically and uniformly searched for transiting signals around these 132 stars in the TESS data and identified five transiting hot super Earths around four stars, including one transiting candidate (TOI-6965.01) with a period of 6\,d around HD~50554. 

Combining TESS photometry, archival RV measurements, and Gaia astrometry, we were able to confirm the planetary nature of the two transiting candidates around HD~50554, even though the outer one with a period of 28\,d was not in our statistical sample. The available RV measurements were only able to constrain the masses of these two super Earths with radii of $1.3\,M_\oplus$ and $1.4\,M_\oplus$ to $<5.3\,M_\oplus$ and $10.4\,M_\oplus$, respectively.

Our statistical sample of 132 stars yields five hot super Earths from four stars. From these we report an enhancement factor of $6.5^{+3.1}_{-2.3}$ in the occurrence rate of hot super Earths by the presence of cold Jupiters. Taking into account the average multiplicity of hot super Earths, we find that the fraction of Sun-like stars with hot super Earths given that there is already at least one cold Jupiter is about 91\%. These numbers indicate a strong correlation between the hot super Earth and cold Jupiter populations.

Below we discuss what our results imply in a broader context.

\section{Implications to the super Earth--cold Jupiter connection} \label{sec:discussion}

\subsection{Different statistical approaches to the conditional probability}

Using RV data, \citet{Rosenthal:2022} previously derived the fraction of stars with close-in small planets conditioned on the presence of outer giants to be $P({\rm I/O})\approx 42\%$. While their different definition of both inner small and outer giant planets is certainly one reason for the relatively low conditional rate, the different statistical approaches are another reason.

{\citet{Rosenthal:2022} introduced an ad hoc statistical framework utilizing a Poisson distribution to model the planet occurrence rate, which is then treated as the conditional probability $P({\rm I|O})$ by imposing a hard upper boundary. To account for planet multiplicity, they only counted the first detected planet in multiple-planet systems. If one is to follow the approach of \citet{Rosenthal:2022} except replacing the Poisson distribution by a binomial distribution, given $N_{\rm det}$ systems with detected HSEs out of $N_\star$ total systems, the posterior distribution of the conditional probability $P({\rm HSE|CJ})$ is given by 
\begin{equation}
    f(p;a,b)=\frac{1}{\mathrm{B}(a,b)}p^{a-1}(1-p)^{b-1},
\end{equation}
where ${\rm B}(a,b)$ is the beta function, $a\equiv N_{\rm det}+1$, $b \equiv N_\star-N_{\rm det}+1$, and
\begin{equation} \label{eqn:neff}
    p \equiv P({\rm HSE|CJ}) \langle p_{{\rm tr}} p_{{\rm det}}\rangle .
\end{equation}
Here the sample completeness is simply a product of the average sensitivity and the transit probability.
\footnote{{In \citet{Zhu:2024}, the completeness correction is applied to $N_\star$, which is not logically correct. While it does not make a difference for the estimation of $P({\rm CJ|SE})$, it does for $P({\rm HSE|CJ})$ because the survey completeness here is very low.}}
The resulting posterior yields $P(\rm HSE|CJ)\approx 45\%$, which is {much} lower than the $P(\rm HSE|CJ)=91\%$ derived from the enhancement factor (Equation \ref{eqn:con_prob}).}

{This large deviation originates from the simplified treatment of sample completeness. With Equation~(\ref{eqn:neff}), it implicitly assumes a uniform underlying distribution of planets. This assumption fails when the parameter space is so broad that the planet intrinsic occurrence rate varies significantly within it, as in our case. Taking the distribution of the intrinsic occurrence rate into account, Equation~(\ref{eqn:neff}) should be corrected to
\begin{equation}
p \equiv P({\rm HSE|CJ}) \cdot \frac{\langle \eta_{{\rm HSE}} p_{{\rm tr}} p_{{\rm det}}\rangle}{\langle \eta_{{\rm HSE}} \rangle}.
\end{equation}
With this correct $p$, the derived $P({\rm HSE|CJ})$ ($\approx 93\%$) becomes consistent with the results estimated through the enhancement factor method.}

{Given the above complications, the conditional probability is actually hard to constrain directly from the sample, whereas the enhancement factor is a more straightforward quantity to constrain. As demonstrated in Section \ref{sec:result}, one can constrain the enhancement factor by directly comparing the number of detected HSEs in CJ systems with the (expected) number around field stars. 
We now apply this approach to the sample of \citet{Rosenthal:2022}
Among the 719 stars in the California Legacy Survey \citep[CLS,][]{Rosenthal:2021}, 43 small planets ($M\sin i < 30\,M_\oplus$) were detected. Applying our definition of a CJ ($0.3 < M_{\rm p}\sin i < 20\,M_{\rm J}$, $P > 300\,\rm days$, and $a < 10\,\rm au$), there are approximately 70 CJs distributed across $\sim60$ stars \citep{Fulton:2021}, using the observed CJ multiplicity of 1.15 for the CLS sample \citep{Li:2026}. Within these CJ systems, there are $\sim13$ small planets ($0.02\text{--}1\,\rm au$, $2\text{--}30\,M_\oplus$). Comparing the ratio of detected small planets in CJ systems (13/60) to the baseline ratio around field stars (43/719) yields an enhancement factor of $\sim4$. This value is significantly larger than the enhancement of $P({\rm I|O})/P(\rm I)\sim1.5$ reported in \citet{Rosenthal:2022}, {again confirming that the statistical approach is, at least partly, responsible for their relatively low conditional rate.}

{We caution that the enhancement factor derived from the CLS survey cannot be directly compared to our result due to different population definitions.
Nevertheless, the reanalysis of the CLS sample, based on a more proper statistical method, corroborates our result that the presence of Cold Jupiter(s) significantly enhances the likelihood of hosting inner small planets.}

\subsection{Impact of stellar metallicity \& mass}

\begin{figure*}[htbp]
    \centering
    \includegraphics[width=\linewidth]{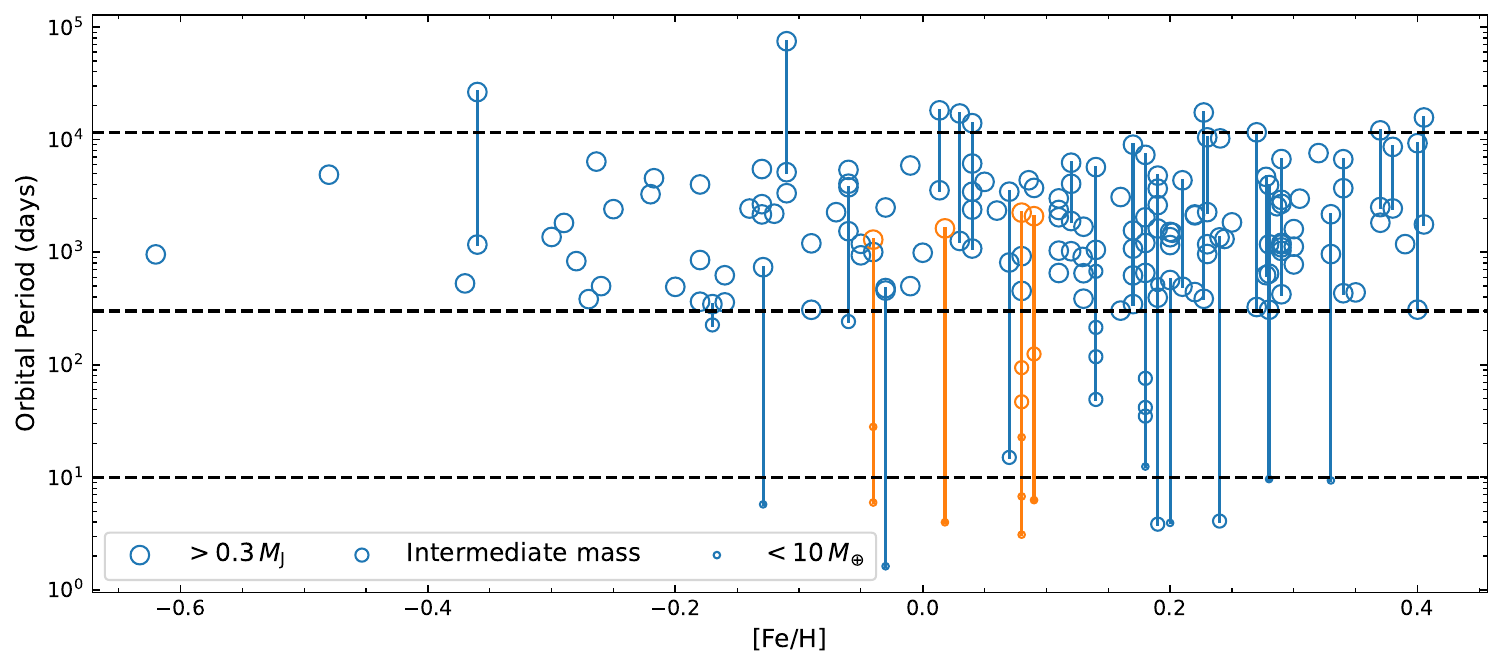}
    \caption{The planetary systems with RV CJs in our sample as a function of host star metallicity. The four systems with transiting HSE discovered by TESS are marked in orange.}
    \label{fig:metal}
\end{figure*}

Recent studies have confirmed that the stellar metallicity plays a role in the correlation. While the positive correlation has been reported around metal-rich stars \citep{Zhu:2024, Bryan:2024, Bonomo:2025}, no consensus has been reached about the metal-poor end. This is probably because those studies all focused on the conditional probability of cold Jupiters given super Earths, $P({\rm CJ|SE})$. As cold Jupiter occurrence correlates strongly with host star metallicity \citep[e.g.,][]{Santos:2001, Santos:2004, Fischer:2005}, the absolute rate of $P({\rm CJ})$ is no more than a few percent if one is limited to sub-solar metallicities, and the value of $P({\rm CJ|SE})$ is probably $\lesssim 10\%$ even if the correlation remains as strong as it is in the metal-rich side. To measure such a small value with high significance and thus confirm (or not) the correlation in the metal-poor side thus requires a large sample size, exceeding what is available at the moment \citep[e.g.,][]{Bryan:2024, Bonomo:2025}.

The derivation of the converted conditional rate, namely $P({\rm SE|CJ})$, is less affected by stellar metallicity, because the occurrence of super Earths correlates much more weakly, if at all, with stellar metallicity \citep[e.g.,][]{Buchhave:2014, Petigura:2018, Zhu:2019}. This approach faces its own challenges, however, as it requires to detect the low-mass super Earths in the presence of cold Jupiters. Although dedicated searches may be able to make it \citep{Delisle:2025}, this requirement is generally difficult for RV method. One way to navigate the challenge would be to combine different detection methods, as is done here. Our sample consists of stars that have cold Jupiter detections from RV, and then we search for inner super Earths with transit. 

If we set the distinction between metal-rich and metal-poor at solar value and exclude one system that has no reliable metallicity measurement, then there are 43 metal-poor and 88 metal-rich CJ systems, and the number of systems with transiting hot super Earths for the two subsamples are one and three, respectively. See Figure~\ref{fig:metal} for an illustration of our sample along the metallicity dimension. The ratios, $1/43$ vs.\ $3/88$, are quite compatible with each other,
\footnote{We note that HIP 54597 has [Fe/H]$\approx-0.22$ \citep{Frensch:2023, Philipot:2023}, although the transiting hot super Earth around it, TOI-6709.01, does not meet our detection threshold.}
and they are both substantially higher, in the statistical sense, than the expected ratio ($\sim0.006$) under the assumption of null correlation (see Section~\ref{sec:result}). Therefore, our results suggest that the super Earth--cold Jupiter correlation remains strong in the metal-poor regime. If we define metal-poor as with [Fe/H]$<0.05$ and metal-rich as with [Fe/H]$>0.05$ to avoid the potential complication of metallicity uncertainties, then we have 34 and 80 in the metal-poor and metal-rich subsamples, and the numbers of hot super Earth detections in TESS are zero and two, respectively. These numbers are still consistent with a positive correlation between super Earths and cold Jupiters across different stellar metallicities, although the alternative explanation is not ruled out. Future studies based on larger samples, in particular by making use of the Gaia giant planet detections \citep{Espinoza-Retamal:2023}, should provide more conclusive results on this matter.

{\citet{Bonomo:2025} and \citet{Bryan:2025} also investigated the dependence of the correlation on stellar mass, finding that the correlation appears only around stars with relatively high masses. For a similar reason to the metallicity dependence, the currently non-detection of the correlation around low-mass stars can simply be explained by the limited CJ detections around such stars. In our sample, the stellar masses of the HSE hosts are broadly consistent with the overall stellar-mass distribution of the full sample, as shown in Figure~\ref{fig:mass_mag}, with no preference for higher stellar masses}. 

\subsection{How about warm super Earths?}

We choose to study the correlation between cold Jupiters and hot super Earths because of the sensitivity limitation of TESS. For super Earths with longer periods (10--300\,d), namely warm super Earths, the transit probability significantly reduces, and the observing strategy of TESS was not able to produce homogeneous period coverage. Nevertheless, at least one of our targets, HD 50554, does contain a transiting warm super Earth (see Section~\ref{sec:validation}). Additionally, as shown in Figure~\ref{fig:sample}, some targets in our sample also host warm super Earths that were detected by RV, and HD 39091 ($\pi$ Mensae) might host a $\sim13\,M_\oplus$ planet at 125-day orbit \citep{Hatzes:2022}. Therefore, (at least some) warm super Earths can get along with cold Jupiters. 

Do warm super Earths also correlate with cold Jupiters, just like their hot siblings? 
From a theoretical perspective, warm super Earths can be easily perturbed by the eccentric outer giants during the dynamical evolution of the system \citep[e.g.,][]{Schlecker:2021, Bitsch:2023}, and the occurrence of warm super Earths does seem to show a negative correlation with increasing metallicity \citep{Petigura:2018}. These would suggest that the occurrence of warm super Earths may be suppressed by the presence of cold Jupiters. 
However, a suppressed occurrence rate of warm super Earths does not necessarily mean a suppressed occurrence rate of warm super Earth systems, and it is the rate of systems that matters more to the connection with cold Jupiters \citep{Zhu:2019}. Unfortunately, this latter quantity cannot be constrained with the currently available data, so we are not yet able to answer the question that is posted in the beginning of this paragraph.

\subsection{Cold super Jupiter vs.\ cold Saturn}
\label{sec:cj-mass}

A recent work by \citet{Lefevre:2025} claimed that the presence of inner low-mass planets ($<0.1\,M_{\rm J}$) preferentially correlates with outer giants of Saturn-mass (0.1--1.5$\,M_{\rm J}$) rather than super-Jupiter ($>1.5\,M_{\rm J}$), based on a re-analysis of the pure RV sample of CLS. While there are several differences in the definitions of close-in planets and distant giants as well as whether restricting to Sun-like stars (FGK dwarfs, as in the present work) or mixing all stellar types and evolution stages (as in \citealt{Lefevre:2025}), here we focus on the main conclusion of that study, namely that super Jupiters do not correlate with close-in low-mass planets.

As shown in Figure~\ref{fig:sample}, the masses (to be more accurate, projected masses) of the cold Jupiters with inner transiting hot super Earths in our sample span almost the entire allowed mass range, from $\approx0.34\,M_{\rm J}$ (HD 219134~h) to $\approx 14\,M_{\rm J}$ (HD 39091~b, or $\pi$ Mensae b). The cold Jupiter in our newly confirmed TESS transiting system, HD 50554~b, also joins the super-Jupiter category that is defined by \citet{Lefevre:2025}, whereas the cold Jupiter in the fourth system (HD 86226~b) would be classified as Saturn-mass. Because the four systems straddle two ends of the cold Jupiter mass, any statistical analysis would be very sensitive to how one divides super-Jupiter vs.\ Saturn-mass. Nevertheless, the existence of systems like $\pi$ Mensae and HD 50554 does seem to contradict the claim of \citet{Lefevre:2025}. In other words, the most massive giant planets can be accompanied by inner super Earths.

The discovery process of the $\pi$ Mensae and HD 50554 systems also shed light on the limitations of pure RV samples. In both systems, the inner super-Earths were not detected by RV first, despite that both stars had received intensive RV observations. Yet $\pi$ Mensae c and HD 50554~c and d have relatively short periods, rendering them relatively high RV amplitudes. The bulk of the super Earth population have {even smaller} RV signals, and disentangling them from the much stronger signals of cold Jupiters---especially eccentric super Jupiters---will be very challenging.

Another interesting feature to note before we conclude this section is that two systems hosting both RV super Jupiters and transiting super Earths are both dynamically hot, with the super Jupiters exhibiting large eccentricities and significant mutual inclinations with respect to the inner super Earths: HD 39091 b has an eccentricity of $\sim0.6$ and an inclination of $\sim54^\circ$, implying a mutual inclination of $\gtrsim36^\circ$ with respect to $\pi$ Men c (\citealt{Feng:2022}; see also \citealt{Xuan:2020, DeRosa:2020, Damasso:2020}); HD 50554 b has an eccentricity of $\sim0.5$ and an inclination of $\sim61^\circ$, corresponding to a mutual inclination of $\gtrsim30^\circ$ relative to the inner transiting planets \citep{Xiao:2023}. The two systems containing Saturn-mass cold giants and inner transiting super Earths do seem to be dynamically colder, as is evidenced by the small eccentricities of both HD 219134~h ($\sim0.06$, e.g.~\citealt{Vogt:2015}) and HD 86226~b ($\sim0.06$, e.g.~\citealt{Teske:2020}). It is unclear whether there is really a dichotomy or a rather smooth distribution that connects the two quite different architectures.

\begin{acknowledgements}
We would like to thank the anonymous referee for comments on the manuscript. This work is supported by the National Natural Science Foundation of China (grant No. 12173021 \& 12133005). We also acknowledge the Tsinghua Astrophysics High-Performance Computing platform for providing computational and data storage resources.
\end{acknowledgements}

\bibliography{sample631}{}
\bibliographystyle{aasjournal}

\end{document}